\documentclass[letterpaper,twocolumn,10pt]{article}
\usepackage{usenix-2020-09}

\usepackage{tikz}
\usepackage{amsmath}
\usepackage{subfigure}
\usepackage{subcaption}
\usepackage{enumitem}
\usepackage{url}
\usepackage{multirow}
\usepackage{comment}
\usepackage{xcolor}
\usepackage{graphicx}
\usepackage{pdfpages}
\usepackage{pifont}
\usepackage{fontawesome}
\usepackage{multicol}
\usepackage[normalem]{ulem}
\usepackage{booktabs}
\usepackage{makecell}
\usepackage{tabularx}
\usepackage{tabulary}

\usepackage{xspace}
\usepackage{algorithm}
\usepackage[noend]{algpseudocode}

\graphicspath{{fig-struct/}{fig-exp/}}
\newcommand{\sysname}{TriVS\xspace}
\newcommand{\algname}{TriSearch\xspace}

\algrenewcommand\textproc[1]{\texttt{#1}}
\algrenewcommand\algorithmicfunction{\textbf{def}}

\newcommand{\para}[1]{\noindent\textbf{#1}}

\begin{document}

\date{}


\title{Look Before You Leap: \\Re-Architecting Disk-Resident Vector Search with \sysname}

\author{
{\rm Yizou Chen$^{1}$ \quad Baotong Lu$^{2}$ \quad Peiqi Yin$^{1}$ \quad Zhisheng Hu$^{1}$ \quad Qi Chen$^{2}$ \quad Ming-Chang Yang$^{1}$}\\[0.5em]
$^{1}$The Chinese University of Hong Kong \quad $^{2}$Microsoft Research\\[0.3em]
{\normalsize\rmfamily\upshape \{chenyz,zshu23,pqyin22,mcyang\}@cse.cuhk.edu.hk \quad \{baotonglu,cheqi\}@microsoft.com}
} %

\maketitle

\begin{abstract}
Disk-resident approximate nearest-neighbor search (ANNS) scales vector search beyond DRAM, yet state-of-the-art systems remain I/O-bound, even with fast SSDs and sophisticated I/O execution.
To understand this bottleneck, we conduct an I/O-centric analysis that breaks each query down into an \textit{approach phase} that locates a query-near region and a \textit{converge phase} that resolves the final top-$k$.
This analysis reveals complementary I/O tradeoffs:
cluster-based search circumvents approach-phase I/O through in-memory routing but reads coarse-grained, full clusters during convergence, whereas graph-based search employs fine-grained, vector-level I/O throughout, enabling efficient convergence at the cost of inevitable I/O during the approach phase.

To combine the complementary I/O benefits of the two granularities, we present \textit{\sysname}, a new paradigm that re-architects disk-resident vector search as a \textit{trilogy: approach coarsely, distill progressively, and converge exactly}.
Its key innovations come in two parts.
First, it decouples the approach and converge phases from a single search granularity, employing cluster-based search to avoid approach-phase I/O and leveraging vector-level I/O for efficient convergence.
Second, to bridge the two granularities, it introduces a new in-memory \textit{distill phase} between the original two, progressively reducing routed clusters to a vector-level shortlist using compact quantized vectors before disk access.
Moreover, coarser clustering shrinks the centroid-routing index, allowing \sysname to reallocate existing memory to distillation rather than expand its memory budget.
Our evaluation on five datasets spanning diverse vector types and scales shows that, at Recall@10=0.9, \sysname achieves on average $3.42\times$ PipeANN's throughput, 51.2\% lower latency, and 63.7\% less disk I/O, while using less memory and storage than PipeANN.
\end{abstract}

\section{Introduction}
Modern search and retrieval systems represent images, text, and other objects as high-dimensional vectors.
Because exact nearest-neighbor search is prohibitively expensive over large high-dimensional collections, approximate nearest-neighbor search (ANNS) trades a small loss in recall for substantially less work~\cite{ANNS_Approximate_motivation}.
As datasets grow to billions of vectors, disk-resident ANNS has become a \textit{de facto} standard for cost-efficient vector search: it offloads full-precision vectors and their large indices (each often requiring hundreds of gigabytes~\cite{cheng2024characterizingdilemmaperformanceindex}) to SSDs while retaining only compact search state in DRAM.
Based on how they organize and access the index and disk-resident data, existing disk-resident ANNS systems generally fall into two categories: 1) \textit{cluster-based ANNS} that manages vectors at a coarse-grained cluster granularity (with SPANN~\cite{SPANN} as a state-of-the-art baseline) and 2) \textit{graph-based ANNS} that organizes vectors into a proximity graph at vector granularity (exemplified by the representative DiskANN~\cite{DiskANN} and PipeANN~\cite{pipeann_osdi}).

While both (cluster-based and graph-based) search paradigms have seen widespread adoption and extensive optimization, we nevertheless observe that even with faster SSDs, they remain inherently I/O-bound, especially under high concurrency.
Specifically, our measurements on a high-end PCIe 5.0 NVMe SSD show that SPANN, DiskANN, and PipeANN stop scaling as query concurrency increases: they all approach the SSD's I/O capacity, causing exposed I/O wait to account for 87--93\% of total query latency at high concurrency.
This bottleneck persists even for PipeANN despite its optimized pipelined execution.
These results identify I/O volume as the primary bottleneck to scalability, establishing a key design principle that scalable disk-resident ANNS systems must pursue: \textbf{\emph{minimize per-query disk I/O}}.

To identify what drives this per-query I/O volume, we break each query down into an \emph{approach phase}, which locates a region likely to contain the nearest neighbors, and a \emph{converge phase}, which examines candidates within that region to determine the final top-$k$.
Our analysis and experiments reveal that \textbf{\emph{fine and coarse search granularities provide complementary I/O benefits across the two phases}}.
Specifically, cluster-based search avoids approach-phase I/O through in-memory centroid routing, but it reads coarse-grained, full clusters during convergence, incurring excessive I/O volume.
By contrast, graph-based search performs fine-grained, vector-level I/O across both phases: this granular retrieval effectively limits the I/O volume during the convergence phase, but it incurs unavoidable I/O overhead during the approach phase.

Aiming at combining the identified, complementary I/O benefits of the two granularities,%
we present \textbf{\emph{\sysname}}, a new paradigm that re-architects disk-resident vector search as a \textbf{\underline{tri}}logy: \textbf{\emph{approach coarsely, distill progressively, and converge exactly}}.
Specifically, \sysname adopts cluster granularity to avoid approach-phase I/O and employs vector granularity for efficient convergence.
Meanwhile, to bridge the huge gap between the two granularities, \sysname introduces an in-memory \emph{distill phase} that progressively reduces routed clusters to a vector-level shortlist using compact quantized vectors before disk access.
This three-phase workflow defines a third path beyond the existing fixed-granularity graph-based or cluster-based approaches: it adaptively leverages progressively-shrinking granularities as the search advances and the query space narrows, establishing a novel search-aware methodology to fundamentally minimize per-query disk I/O.

Making this design practical requires balancing I/O reduction with in-memory compute overhead.
To meet this requirement, we develop \algname, a vector search algorithm that progressively increases quantization precision as the candidate set shrinks.
It combines the progressive refinement with early termination and lower-bound pruning to limit in-memory computation and exact-vector I/O.
In addition, \sysname reallocates rather than expands memory.
By enabling coarser clustering, in-memory distillation reduces the number of centroids and shrinks the routing index, freeing DRAM for the quantized vectors needed by the distill phase.

We comprehensively evaluate \sysname across datasets spanning diverse vector types and scales.
Across five datasets at Recall@10=0.9, \sysname achieves, on average, $3.42\times$ PipeANN's throughput, 51.2\% lower latency, and 63.7\% less disk I/O, while using less memory and storage than PipeANN.

This paper makes three contributions:

\begin{itemize}[leftmargin=*,nosep]
  \item We present an I/O-centric analysis that identifies disk I/O as the scalability bottleneck in existing disk-resident ANNS and reveals how fine and coarse search granularities shift I/O between the approach and converge phases (\S\ref{sec:motivation}).
  \item We design \sysname as a trilogy that decouples search granularity through in-memory distillation, and develop the \algname algorithm and an efficient implementation to realize this design (\S\ref{sec:design}).
  \item We comprehensively evaluate \sysname across diverse workloads and scales, showing that it achieves higher throughput and lower latency than state-of-the-art disk-resident ANNS systems by reducing disk I/O (\S\ref{sec:eval}).
\end{itemize}

\section{Background}

\subsection{Disk-Resident ANNS}~\label{sec:bg:DiskANNS}
Given a vector dataset $X$ and a query vector $q$, vector search retrieves the $k$ vectors in $X$ closest to $q$ under a chosen distance metric.
Because exact search is prohibitively expensive on high-dimensional datasets~\cite{ANNS_Approximate_motivation}, modern vector search systems~\cite{facebook_web_search, youtube_recommentation, RAG_nips_2020} rely on approximate nearest-neighbor search (ANNS) to significantly reduce search cost at the expense of a small, acceptable accuracy loss. 
ANNS accuracy is typically measured by Recall@$k$, the fraction of true top-$k$ neighbors retrieved.
As datasets grow to hundreds of millions or billions of high-dimensional vectors, the full-precision vectors alone can occupy hundreds of gigabytes, and the index can require comparable or greater capacity~\cite{DiskANN,FlowANN,cheng2024characterizingdilemmaperformanceindex}.
Large-scale vector search therefore commonly uses disk-resident ANNS, storing the bulk of the vectors and index on SSD while retaining compact search state in DRAM~\cite{DiskANN,SPANN,Starling,pipeann_osdi,Helmsman}.

\begin{figure*}[t]
  \centering
  \begin{minipage}[t]{0.425\textwidth}
    \centering
    \includegraphics[width=\linewidth]{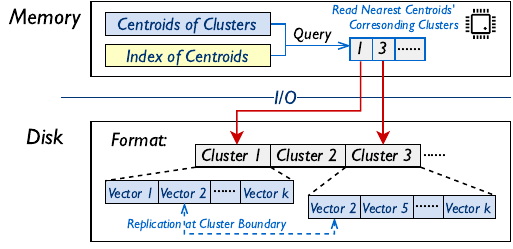}
    {\small\textbf{(a) Cluster-based ANNS:} \textit{organized at cluster granularity}}
  \end{minipage}
  \hfill
  \begin{minipage}[t]{0.425\textwidth}
    \centering
    \includegraphics[width=\linewidth]{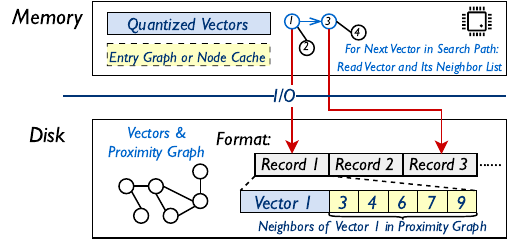}
    {\small\textbf{(b) Graph-based ANNS:} \textit{organized at vector granularity}}
  \end{minipage}
  \captionsetup{skip=6pt}
  \caption{\textbf{Data placement and query flow in existing disk-resident ANNS.}
  \textit{
  Cluster-based systems locate nearby clusters through an in-memory centroid index and read each selected cluster as a unit, avoiding navigation I/O but potentially fetching many vectors in each cluster.
  Graph-based systems store a full-precision vector and its proximity-graph neighbors in each disk record, so expanding a vertex (i.e., vector) requires I/O to read its neighbor list, which guides the subsequent search path.
  }}
  \label{fig:baseline-overviews}
\end{figure*}

Existing disk-resident ANNS systems can be classified by the granularity at which they organize the index and disk-resident data.
Cluster-based systems such as SPANN~\cite{SPANN} use coarse-grained cluster granularity, whereas graph-based systems such as DiskANN~\cite{DiskANN} and PipeANN~\cite{pipeann_osdi} use fine-grained vector granularity.
The chosen granularity shapes index size, data placement, query workflow, and disk access, as Figure~\ref{fig:baseline-overviews} summarizes.

\para{Cluster-based Disk-Resident ANNS.}
Cluster-based systems such as SPANN~\cite{SPANN} organize vectors at cluster granularity: nearby vectors are assigned to a cluster, stored together on disk, and represented by its centroid (Figure~\ref{fig:baseline-overviews}(a)).
Because one centroid represents many vectors, the routing index contains far fewer entries than a per-vector graph and can remain in memory.
At query time, the system searches the in-memory index for the centroids closest to the query, then reads the corresponding clusters from disk in a batch and computes the exact distance to each retrieved vector.
This coarser organization eliminates navigation I/O and enables batched reads, but fetching at cluster granularity can amplify disk I/O by transferring every vector in each selected cluster.
To maintain recall while limiting cluster probes, SPANN~\cite{SPANN} replicates boundary vectors across neighboring clusters, trading storage overhead for lower per-query I/O.

\para{Graph-based Disk-Resident ANNS.}
Graph-based systems such as DiskANN~\cite{DiskANN} and PipeANN~\cite{pipeann_osdi} organize vectors into a proximity graph at vector granularity: each vertex represents a vector, and edges connect nearby vectors (Figure~\ref{fig:baseline-overviews}(b)).
Because each vector links to multiple neighbors, the graph edges can collectively rival the raw vector data in size~\cite{FlowANN,cheng2024characterizingdilemmaperformanceindex}.
Under practical DRAM budgets, graph-based systems keep compact quantized vectors in memory and co-locate each exact vector with its graph neighbor list in one disk record, so one read returns both the full-precision vector and graph edges needed for exact distance computation and continued traversal.
These systems commonly use best-first search~\cite{Graph_Based_Vector_Search_Evaluation_SIGMOD,Survey_Graph_Based_Vector_Search_VLDB}, which repeatedly expands the closest unvisited vector in a candidate list. 
Each expansion reads the exact vector and its neighbor list from disk, computes its exact distance to the query, and uses memory-resident quantized vectors to score its neighbors and update the candidate list.
Compared with cluster-based approaches, this vector-granularity organization keeps individual reads selective and reduces redundant data access for high system throughput, but may incur higher query latency due to the dependent I/O inherent in graph traversal.
The state-of-the-art PipeANN~\cite{pipeann_osdi} pipelines reads across search steps and adapts I/O parallelism to overlap I/O with computation, but retains the same per-vector disk layout and access pattern.

\subsection{Vector Quantization}
Vector quantization maps a high-dimensional vector to a compact code that approximates its distance to a query.
Common methods include scalar quantization (SQ), product quantization (PQ)~\cite{PQ}, OPQ~\cite{OPQ}, LVQ~\cite{LVQ_VLDB}, and RaBitQ~\cite{RaBitQ,Extended_RaBitQ}.
For high-dimensional vectors, SAQ~\cite{SAQ_SIGMOD} segments PCA-projected vectors and allocates more bits to high-impact dimensions, while DADE~\cite{DADE_VLDB}, ADSampling~\cite{ADSampling_SIGMOD}, and LeanVec~\cite{LeanVec} reduce or adapt the dimensions used for distance evaluation.
Across these methods, different code widths, quantizer configurations, and dimension subsets provide different trade-offs among accuracy, memory, and computation.
Disk-resident graph systems such as DiskANN~\cite{DiskANN} and PipeANN~\cite{pipeann_osdi} keep PQ codes in memory to guide vector-level navigation.
After reading an expanded vector's neighbor list from disk, the system uses quantized codes to estimate each neighbor's distance to the query and updates the ordered candidate list which it then uses to select the next graph records.
This in-memory distance estimation avoids fetching every neighbor's full-precision vector from disk.

Among these methods, RaBitQ~\cite{RaBitQ,Extended_RaBitQ} is a prominent state-of-the-art quantizer that provides multiple precision levels and theoretical error bounds.
Given a reference center $\mathbf{c}$, typically a cluster centroid, RaBitQ normalizes the residuals of the data vector and query, applies a shared random rotation, and represents each residual dimension with one or more bits.
A one-bit code provides an initial estimate, while additional bit planes progressively refine it and reuse the earlier computation.
During index construction, RaBitQ stores each vector's code together with a few precomputed terms derived from that vector and its center.
At precision level $\ell$, search evaluates the code against the transformed query residual and combines the result with these terms to obtain an estimated distance $D_\ell$.
RaBitQ also uses a stored error term and the query residual norm to derive an error width $E_\ell$, which bounds the exact distance within $[D_\ell-E_\ell,D_\ell+E_\ell]$ at the specified confidence level.
We denote the corresponding lower bound as $\mathrm{LB}_\ell=D_\ell-E_\ell$.

\section[Observation and Motivation: An I/O-Centric View of Disk-Resident ANNS]{Observation and Motivation:\ An I/O-Centric View of Disk-Resident ANNS}
\label{sec:motivation}

In this section, we take an I/O-centric view of disk-resident ANNS.
We first show that existing systems scale poorly as query concurrency increases, even on a high-end PCIe 5.0 NVMe SSD, and identify disk I/O as their shared bottleneck.
To understand where this disk I/O arises, we break down the vector search process into approach and converge phases and examine the granularity at which each phase operates.
This analysis reveals that existing systems apply one granularity throughout the vector search process, thereby determining the amount and distribution of disk I/O across its phases.

\subsection{Existing Systems Remain I/O-Bound}\label{sec:motivation:scalability}

\begin{figure}[t]
  \centering
  \includegraphics[width=\columnwidth]{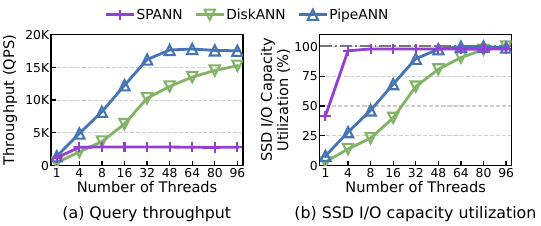}
  \captionsetup{skip=5pt}
  \caption{\textbf{Throughput scaling and SSD I/O utilization.}
  \textit{On DEEP-100M at Recall@10=0.9, each system stops scaling as it approaches the SSD I/O capacity.}}
  \label{fig:baseline-qps-scaling}
\end{figure}

We begin by measuring the throughput scalability of SPANN, DiskANN, and PipeANN on DEEP-100M at Recall@10=0.9, using 1--96 query threads on the same PCIe 5.0 NVMe SSD.
Figure~\ref{fig:baseline-qps-scaling}(a) shows that all three systems scale sublinearly: SPANN reaches its throughput plateau at only 4 threads, while DiskANN and PipeANN increasingly fall below their early scaling trends.
Figure~\ref{fig:baseline-qps-scaling}(b) links this scaling behavior to SSD I/O utilization.
SPANN reaches 96\% of the SSD I/O capacity~\footnote{We use SSD I/O capacity to refer to the maximum sustainable read IOPS at each system's access granularity, measured using fio~\cite{fio}.} at only 4 threads, where its throughput plateaus.
As concurrency increases, DiskANN and PipeANN also approach the capacity and deliver diminishing throughput gains.
Thus, even on a high-end SSD, existing systems cannot sustain throughput growth as query concurrency drives SSD I/O utilization toward capacity.

\begin{figure}[t]
  \centering
  \includegraphics[width=\columnwidth]{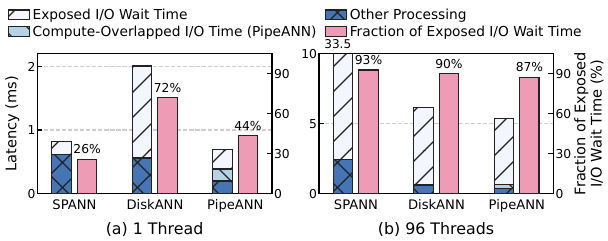}
  \captionsetup{skip=5pt}
  \caption{\textbf{Latency decomposition and exposed I/O wait on DEEP-100M at Recall@10=0.9.}
  \textit{
  At 1 thread, SPANN exposes less I/O wait because it avoids the dependent I/O rounds required to traverse a disk-resident proximity graph; at full system load (96 threads), however, I/O wait dominates latency in all three systems.}}
  \label{fig:baseline-latency-io-breakdown}
\end{figure}

To examine how this I/O bottleneck manifests in per-query latency, we decompose latency and quantify the disk wait time exposed to each query in Figure~\ref{fig:baseline-latency-io-breakdown}.
At 1 thread, SPANN's exposed I/O wait time constitutes 26\% of latency, less than in DiskANN and PipeANN because it can batch cluster reads and avoid dependent I/O rounds in graph traversal.
Although PipeANN optimizes I/O execution to overlap part of the I/O time with computation, its exposed I/O wait time remains significant at 44\% of query latency.
Moreover, at full system load (96 threads), all three systems are I/O-dominated, with the exposed I/O wait time accounting for 87--93\% of latency.
These measurements identify I/O volume as the primary scalability limit, with dependent I/O rounds imposing an additional latency cost at low load.

In summary, existing systems, both cluster- and graph-based, stop scaling as they approach SSD I/O capacity.
Optimizations such as compute--I/O overlap (e.g., PipeANN~\cite{pipeann_osdi})
can hide part of the I/O latency at low concurrency, but they leave per-query I/O volume unchanged.
Consequently, at high concurrency, latency and throughput remain bounded by the SSD's I/O capacity.
Scalable disk-resident ANNS must therefore follow a principle: \textbf{\emph{minimize per-query disk I/O}}.

\begin{figure*}[t]
  \centering
  \begin{minipage}[t]{0.49\textwidth}
    \centering
    \includegraphics[width=\linewidth]{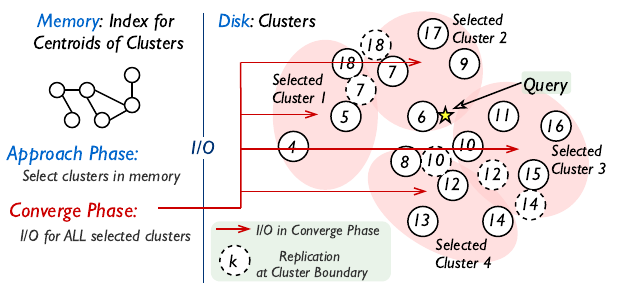}
    {\small\textbf{(a) Cluster-based ANNS:} \textit{cluster-level approach \& converge phases}}
  \end{minipage}
  \hfill
  \begin{minipage}[t]{0.49\textwidth}
    \centering
    \includegraphics[width=\linewidth]{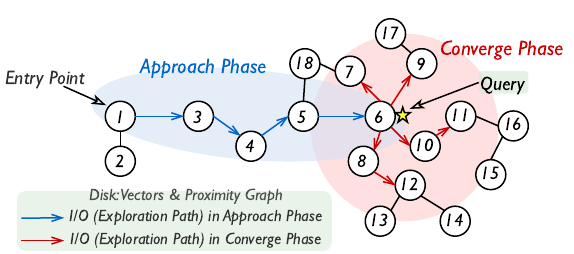}
    {\small\textbf{(b) Graph-based ANNS:} \textit{vector-level approach \& converge phases}}
  \end{minipage}
  \captionsetup{skip=5pt}
  \caption{\textbf{Existing disk-resident ANNS systems apply one granularity throughout the search process.}
  \textit{Cluster-based systems select query-near clusters in memory and read every vector in them, with boundary replication further amplifying the transfer. Graph-based systems follow a vector-level path toward the query and expand individual graph records around it.}}
  \label{fig:phase-granularity}
\end{figure*}

\begin{figure}[t]
  \centering
  \includegraphics[width=0.9\columnwidth]{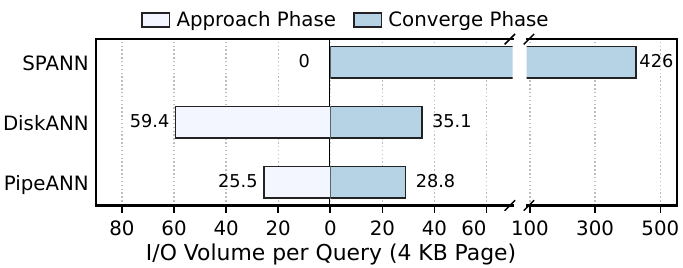}
  \captionsetup{skip=5pt}
  \caption{\textbf{Existing architectures incur phase-specific I/O.}
  \textit{DEEP-100M at Recall@10=0.9.
  Cluster-based SPANN eliminates approach-phase I/O but transfers far more data during the converge phase.
  Graph-based DiskANN and PipeANN limit converge-phase I/O but incur approach-phase I/O.}}
  \label{fig:graph-approach-converge-io}
\end{figure}

\subsection{Different Search Granularities Determine Disk I/O}\label{sec:motivation:granularity}

Having established disk I/O as the shared bottleneck, we next examine how it arises during vector search.
Following prior analyses of vector-index traversal~\cite{VBASE_OSDI,pipeann_osdi,FlowANN}, we break each query down into two analytical phases: the \emph{approach phase} and the \emph{converge phase}.
The approach phase moves the search from its entry point into a region likely to contain the query's nearest neighbors.
The converge phase examines candidates within that region to determine the final top-$k$.
We apply this analytical lens to cluster- and graph-based vector search and study the disk I/O incurred in each phase.

Our analysis and experiments reveal a common source of excessive I/O: \textit{\textbf{existing architectures couple both phases to a single search granularity, optimizing one phase at the expense of the other.}}
This mismatch arises because the two phases favor different granularities: cluster-level routing keeps the index compact enough to reside in memory, eliminating approach-phase I/O, whereas vector-level access restricts converge-phase I/O to selected vectors.

\para{Cluster-based disk-resident ANNS eliminates approach-phase I/O at the cost of amplified converge-phase I/O.}
SPANN~\cite{SPANN} applies cluster granularity to both phases, as Figure~\ref{fig:phase-granularity}(a) illustrates.
During the approach phase, its compact centroid index selects query-near clusters entirely in DRAM, avoiding routing I/O.
However, the converge phase fetches each selected cluster in full, although only a small fraction of its vectors can contribute to the final result.
Figure~\ref{fig:graph-approach-converge-io} quantifies this tradeoff: SPANN incurs no approach-phase I/O but transfers 426 pages per query during convergence, 7.8$\times$ PipeANN's total I/O.

\para{Graph-based disk-resident ANNS reduces converge-phase I/O through selective vector reads but incurs dependent approach-phase I/O.}
DiskANN~\cite{DiskANN} and PipeANN~\cite{pipeann_osdi} operate at vector granularity in both phases, treating each vector as a vertex in the proximity graph (Figure~\ref{fig:phase-granularity}(b)).
During the approach phase, best-first search follows a vector-level path from the entry point toward the query; during the converge phase, it continues expanding individual nodes around the query.
The full vector-level navigation state includes one graph record and adjacency list per vector and is too large to retain entirely in DRAM under a practical serving budget, so each expansion reads a disk record containing a full-precision vector and its adjacency list.
This fine-grained access is selective during the converge phase because the search fetches individual vectors rather than the surrounding region.
During the approach phase, however, each expansion must read the current vector's disk-resident neighbor list before discovering new candidates, introducing dependent graph-record I/O along the navigation path.
As shown in Figure~\ref{fig:graph-approach-converge-io}, DiskANN and PipeANN respectively read 59.4 and 25.5 pages per query in approach phase, followed by 35.1 and 28.8 pages in converge phase~\footnote{For each query, we use completion of exact evaluation for the closest vector in its final top-$k$ (e.g., top-10) as the phase boundary; pages completed by then are attributed to approach phase, and later pages to converge phase.}.

In summary, our analysis traces substantial I/O in existing systems to coupling both phases to a single search granularity.
Cluster- and vector-level search granularities each impose phase-specific I/O costs, yet provide complementary I/O benefits for the approach and converge phases, respectively.

\section{\sysname}
\label{sec:design}

To reduce I/O costs both in approach and converge phases, we present \textit{\textbf{\sysname}}, which decouples their search granularities by assigning cluster granularity to the approach phase and vector granularity to the converge phase, then bridges the two phases through an in-memory distill phase that narrows promising clusters to a vector-level shortlist before exact-vector I/O.
\sysname thereby organizes disk-resident vector search as a \textit{\textbf{Tri}logy}: approach coarsely, distill progressively, and converge exactly.
\S\ref{sec:distillation} derives the granularity decoupling and its in-memory enabler; \S\ref{sec:index_org} maps the design onto index construction and data layout; and \S\ref{sec:trisearch} presents the query workflow and the \algname algorithm, which carries each query through the trilogy while bounding both in-memory computation and disk I/O.
\S\ref{sec:optimizations} details the optimizations that make \sysname more efficient, while \S\ref{sec:design-discussion} discusses support for online updates.

\subsection{Decoupling Search Granularity with In-Memory Distillation}
\label{sec:distillation}

Decoupling search granularity could capture the I/O benefits of both existing paradigms: a cluster-level approach phase avoids vector-level graph-traversal I/O, while a vector-level converge phase preserves selective reads rather than fetching full clusters.
However, these granularities cannot be composed directly: the cluster-level approach phase identifies promising clusters, while the vector-level converge phase must select a small subset of the vectors within them.
Without an in-memory mechanism to select this subset, the system must read and scan the selected clusters in full, recreating cluster-level convergence and its coarse-grained I/O.

We observe that compact quantized vectors, which can be kept in memory, provide the missing bridge between cluster-level approach and vector-level convergence.
Even within query-near clusters, most vectors remain sufficiently far from the query to be pruned using approximate distance estimates computed from these in-memory representations.
After cluster routing, \sysname can therefore scan the quantized vectors entirely in memory, discard unlikely candidates, and retain only a vector-level shortlist for the converge phase.

\begin{figure}[t]
  \centering
  \includegraphics[width=\columnwidth]{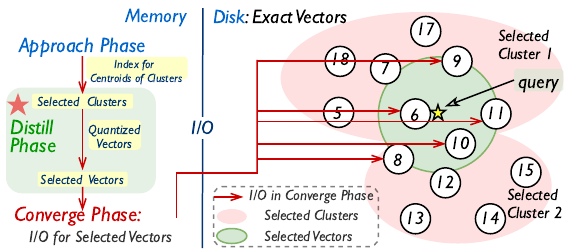}
  \caption{\textbf{\sysname decouples search granularity through in-memory distillation.}
  \textit{The approach phase selects clusters using an in-memory centroid index; the distill phase scans their quantized vectors and emits a vector-level shortlist; the converge phase reads only the shortlisted exact vectors from disk rather than every vector in the selected clusters.}}
  \label{fig:triann-phases}
\end{figure}

\begin{figure*}[t]
  \centering
  \includegraphics[width=0.85\textwidth]{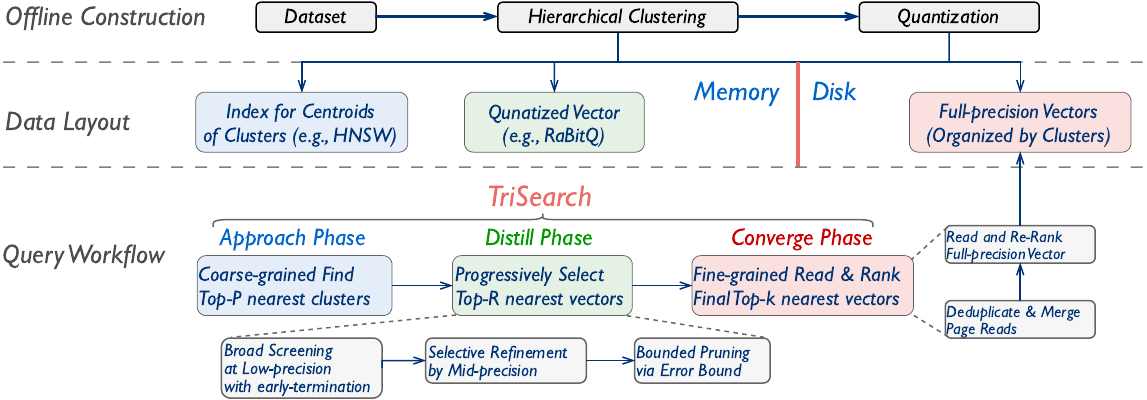}
  \caption{\textbf{\sysname overview.}
  Offline construction creates size-bounded clusters, assigns each vector to one cluster, and materializes a centroid index and cluster-organized quantized vectors in memory alongside cluster-organized full-precision vectors on disk.
  During search, the \algname algorithm routes to query-near clusters, distills their quantized vectors to a vector-level shortlist entirely in memory, and only then reads and re-ranks the shortlisted full-precision vectors.}
  \label{fig:triann-design}
\end{figure*}

This in-memory reduction closes the granularity gap between the cluster-level approach phase and the vector-level converge phase; we call it the \emph{distill phase}.
The \emph{approach}, \emph{distill}, and \emph{converge} phases together define \sysname, a third paradigm for disk-resident ANNS beyond conventional graph- and cluster-based designs.
As shown in Figure~\ref{fig:triann-phases}, \sysname organizes vector search as a trilogy: it approaches query-near regions at cluster granularity, distills the selected clusters into a vector-level shortlist entirely in memory, and converges over only the shortlisted exact vectors on disk.
Compared with existing graph-based and cluster-based workflows (Figures~\ref{fig:phase-granularity}(a) and~\ref{fig:phase-granularity}(b)), \sysname avoids vector-level graph-traversal I/O in the approach phase without resorting to full-cluster reads in the converge phase.
The distill phase provides the in-memory bridge between these granularities, transforming promising clusters into individual candidates before exact-vector I/O.

\para{Distillation reallocates rather than expands memory.}
A natural concern is whether using in-memory quantized vectors for distillation increases memory usage.
It does not: \sysname reallocates existing memory to support distillation rather than expanding its total memory footprint.
On the one hand, compared with graph-based systems, distillation adds no new form of per-vector evidence: these systems already keep compact quantized vectors in memory to guide search.
PipeANN~\cite{pipeann_osdi} pairs these codes with a sampled entry graph, while \sysname pairs them with a centroid index of comparable size.
On the other hand, compared with cluster-based SPANN~\cite{SPANN}, \sysname can scan more and larger clusters because evaluating their in-memory quantized codes does not incur additional full-cluster disk reads.
This flexibility elegantly enables coarser clustering and removes the need for boundary replication.
Coarser clustering produces fewer clusters and therefore fewer centroids, shrinking the centroid-routing index and making room for one compact code per vector.
Assigning each vector to one cluster further prevents those codes from being replicated.
Together, these reallocations let distillation change the search workflow without increasing memory usage.
Our evaluation in \S\ref{sec:eval:overall} confirms that \sysname uses less memory than PipeANN at both 100M and 1B scales.

\subsection{Index Organization}
\label{sec:index_org}

With in-memory distillation providing the missing cluster-to-vector interface, \sysname separates its serving state by phase: memory-resident structures support the approach and distill phases, while full-precision vectors remain on disk until distillation finalizes the shortlist.
Figure~\ref{fig:triann-design} shows how offline construction produces this layout and how queries use it at serving time.

\para{Offline index construction.}
Given a vector dataset, \sysname first applies size-bounded hierarchical $k$-means with a target cluster size $T$ (128 vectors in our implementation), as shown at the top of Figure~\ref{fig:triann-design}.
Specifically, it recursively applies $k$-means to any cluster containing more than $2T$ vectors.
\sysname deliberately adopts much coarser clustering: its cluster count is only about 0.7--0.8\% of the vector count, compared with roughly 10--12\% for SPANN~\cite{SPANN}.
This coarser clustering is both practical and beneficial: in-memory distillation can scan larger clusters without full-cluster disk reads, while the smaller centroid index leaves more memory for quantized vectors.
Notably, in \sysname, each vector belongs to exactly one cluster, avoiding boundary replication and thereby reducing memory and storage overhead while simplifying construction.
For datasets that exceed available construction memory, \sysname processes the hierarchy one subtree at a time.

After clustering, \sysname builds an in-memory routing index (e.g., HNSW~\cite{HNSW}) over the final cluster centroids.
Within each cluster, it quantizes vectors relative to the cluster centroid using RaBitQ~\cite{RaBitQ}.
The per-cluster reference reduces residual spread and improves distance estimation during the distill phase.
Finally, \sysname stores the quantized vectors contiguously by cluster in memory and packs the full-precision vectors in the same cluster order on disk.

\para{Serving-time data layout.}
At serving time, \sysname keeps the centroid index, vector IDs, exact-vector locations, and RaBitQ codes in memory, with per-vector state laid out contiguously by cluster.
The cluster-contiguous RaBitQ representation supports incremental refinement while occupying only two bits per dimension.
Consequently, the approach and distill phases access only DRAM; disk reads begin only after distillation finalizes the shortlist and resolves each candidate's exact-vector location.

\subsection{The \algname Algorithm}
\label{sec:trisearch}

We design \algname, a vector search algorithm that carries each query through \sysname's three-phase workflow: approach, distill, and converge.
Across these phases, \algname bounds both in-memory computation and exact-vector I/O.
First, the cluster-level approach phase searches the in-memory centroid index and returns up to $P$ query-near clusters.
Next, the distill phase progressively filters vectors in the selected clusters using cluster-contiguous RaBitQ codes entirely in memory, forming a shortlist of at most $R$ candidate vectors and their disk locations.
Finally, the vector-level converge phase reads only the shortlisted full-precision vectors from disk and re-ranks them to return the final top-$k$.
Because distillation identifies all exact-vector accesses before disk I/O begins, these reads can be issued as a batch, reducing latency by avoiding dependent I/O rounds.
Figure~\ref{fig:triann-design} provides an overview of this three-phase workflow; Algorithm~\ref{alg:trisearch} details its complete execution.

\begin{algorithm}[!t]
  \caption{Pseudocode for \algname.}
  \label{alg:trisearch}
  \small
  \begin{algorithmic}[1]
    \Statex \textbf{Input:} For query $q$, return its top-$k$ nearest neighbors;
    \Statex \hspace{\algorithmicindent}$P$: maximum clusters probed;
    \Statex \hspace{\algorithmicindent}$R$: maximum exact vectors re-ranked (I/O);
    \Statex At precision level $\ell$, $D_\ell$ and $\mathrm{LB}_\ell$ denote the estimated distance and a lower bound on the distance, respectively.
    \Statex \vspace{-0.8em}
    \Function{\algname}{$q,k;P,R$}
      \Statex \textcolor{gray}{\textit{// 1. Cluster-level approach phase}}
      \State $C \gets \Call{RouteClusters}{q,P}$ \label{alg:route_clusters}
      \State $H \gets$ empty max heap of capacity $R$, keyed by $D_{\mathrm{mid}}$ \label{alg:init_heap}
      \State $W \gets$ early-termination window derived from $P$

      \Statex \textcolor{gray}{\textit{// 2. In-memory distill phase}}
      \Statex \textcolor{gray}{\textit{// Broad screening, selective refinement, and ranking}}
      \ForAll{$c \in C$ in routing order} \label{alg:scan_begin}
        \State $updated \gets \Call{ProgressiveScan}{q,c,H}$ \label{alg:progressive_scan}
        \If{$updated=\mathrm{false}$ for $W$ consecutive clusters} \label{alg:stable_stop}
          \State \textbf{break}
        \EndIf
      \EndFor

      \Statex \textcolor{gray}{\textit{// Exact-access selection}}
      \State $S \gets$ the top-$k$ candidates in $H$ \label{alg:topk_seed}
      \State $\tau_k \gets$ the $k$th-smallest $D_{\mathrm{mid}}$ in $H$
      \State $S \gets S \cup \{x \in H: \mathrm{LB}_{\mathrm{mid}}(q,x) \leq \tau_k\}$ \label{alg:mid_filter}
      \State $S \gets \Call{AddSamePageCandidates}{S,H}$ \label{alg:page_closure}

      \Statex \textcolor{gray}{\textit{// 3. Vector-level converge phase}}
      \State $A \gets \Call{PlanPages}{S}$ \label{alg:plan_pages}
      \State \Return $\Call{ExactRerank}{q,S,A,k}$ \label{alg:exact_rerank}
    \EndFunction

    \Statex \vspace{-0.8em}
    \Function{ProgressiveScan}{$q,c,H$} \label{alg:progressive_begin}
      \State $updated \gets \mathrm{false}$
      \ForAll{vector blocks $b$ in cluster $c$} \label{alg:block_scan}
        \State $(D_{\mathrm{low}},\mathrm{LB}_{\mathrm{low}}) \gets \Call{EvaluateLow}{q,b}$ \label{alg:low_eval}
        \ForAll{$x \in b$} \label{alg:candidate_loop}
          \If{$\mathrm{LB}_{\mathrm{low}}(q,x)<\Call{Threshold}{H}$} \label{alg:low_filter}
            \State $(D_{\mathrm{mid}}[x],\mathrm{LB}_{\mathrm{mid}}[x]) \gets \Call{EvaluateMid}{q,x}$ \label{alg:mid_refine}
            \If{\Call{UpdateHeap}{$H,x,D_{\mathrm{mid}}[x]$}} \label{alg:update_heap}
              \State $updated \gets \mathrm{true}$
            \EndIf
          \EndIf
        \EndFor
      \EndFor
      \State \Return $updated$
    \EndFunction
  \end{algorithmic}
\end{algorithm}

\para{Cluster-level approach phase.}
Given a query $q$, \algname first searches the DRAM-resident centroid index using standard HNSW~\cite{HNSW} to locate query-near clusters (Algorithm~\ref{alg:trisearch}, Line~\ref{alg:route_clusters}).
The search returns an ordered set $C$ of at most $P$ clusters, where $P$ is the cluster-probe budget, and uses a search width of $ef=1.2P$.
These routed clusters become the input to in-memory distillation.
At this point, the query has accessed only centroids; no disk-resident vector has been read.

\para{Progressive in-memory distillation.}
After the approach phase, the routed clusters still contain a broad set of vectors.
Granularity decoupling shifts cluster-to-vector candidate selection from disk reads to quantized-distance evaluation over the routed vectors in DRAM.
This additional computation must remain small enough to avoid replacing the I/O bottleneck with a CPU bottleneck.
Evaluating every routed vector with a more accurate quantized representation would be too costly, whereas using only a coarse representation would leave too many candidates for exact-vector I/O.
To balance these costs, \algname progressively increases quantization precision as the candidate set shrinks: it first screens every routed vector using 1-bit RaBitQ, then reads one additional bit plane only for competitive candidates to evaluate them at mid precision (2-bit RaBitQ).

To implement this progression, \algname visits the routed clusters in order and calls \texttt{ProgressiveScan} on their compact vector codes (Lines~\ref{alg:scan_begin}--\ref{alg:progressive_scan}).
For every vector block, the 1-bit RaBitQ evaluation computes a low-precision distance estimate for each vector (Lines~\ref{alg:block_scan}--\ref{alg:low_eval}).
\algname maintains an $R$-entry max heap $H$ ordered by mid-precision distance (2-bit RaBitQ).
The capacity $R$ bounds the refined candidate pool passed to exact-access selection.
Notably, RaBitQ pairs each estimate with a confidence-controlled lower bound, which \algname uses to decide whether the vector merits 2-bit refinement.
Before $H$ is full, every vector remains eligible for refinement; afterward, the worst mid-precision distance in $H$ becomes the admission threshold.
\algname computes the mid-precision distance for a vector only when its 1-bit RaBitQ lower bound falls below this threshold, then uses the refined distance to update the $R$-entry heap $H$ if it enters the current top $R$ (Lines~\ref{alg:low_filter}--\ref{alg:update_heap}).

Every successful refinement tightens the heap threshold for subsequent vectors, making later screening more selective.
Across clusters, the absence of heap updates provides an early-termination signal: \algname stops after $W$ consecutive routed clusters fail to update $H$, where $W$ is derived from $P$ (Lines~\ref{alg:scan_begin}--\ref{alg:stable_stop}).
Together, within-cluster promotion and cluster-level early termination reserve increasingly precise computation for increasingly smaller candidate sets.

\para{Exact-access selection.}
When the scan terminates, $H$ contains at most $R$ candidates with mid-precision distances and lower bounds.
\algname seeds the exact-access shortlist $S$ with the approximate top-$k$ and uses their $k$th-best distance as a threshold for the remaining candidates.
\algname then retains any remaining candidate whose mid-precision lower bound does not exceed this threshold and adds candidates stored on pages already selected for $S$, since their exact evaluation requires no additional page read (Lines~\ref{alg:topk_seed}--\ref{alg:page_closure}).

\para{Vector-level converge phase (exact re-ranking).}
After the shortlist $S$ is finalized, \algname maps its vectors to disk pages, deduplicates and coalesces the reads, and computes exact distances from the full-precision vectors as the requested pages arrive (Lines~\ref{alg:plan_pages}--\ref{alg:exact_rerank}).
It re-ranks the shortlist by these exact distances and returns the top-$k$ vectors.

\para{Generality across quantizers.}
\algname is not tied to RaBitQ.
Its low- and mid-precision stages can use different code widths, quantizer configurations, or dimension subsets from different quantization methods.
\algname uses RaBitQ's theoretical lower bounds to enhance candidate promotion and pruning, but quantizers without such bounds can rely on estimated distances, fixed refinement ratios, or quantizer-specific pruning criteria.
Different quantizers may affect accuracy and efficiency but preserve \algname's three-phase workflow.

\subsection{Implementation and Optimizations}\label{sec:optimizations}

\algname bounds the work performed in each phase of the search process; we further optimize the complete workflow to reduce computation, I/O, and memory footprint.

\para{Efficient progressive distillation.}
For distillation, each cluster stores a 1-bit-per-dimension RaBitQ base code plus one additional bit plane, forming a nested 2-bit-per-dimension representation.
The codes are arranged in 32-vector blocks for an AVX-512 FastScan kernel~\cite{Faiss}.
The kernel computes the base-precision estimate and lower bound together; refinement reuses this work and evaluates only the additional bit plane.
The final partial block stores only valid lanes and is reconstructed in thread-local scratch space for aligned access, avoiding per-cluster padding.

\para{Compact in-memory state.}
The centroid HNSW stores 8-bit scalar-quantized (SQ8) centroids, a configuration supported by Faiss~\cite{Faiss} and suited to identifying a broad set of promising clusters.
Within each cluster, \sysname also encodes the per-vector RaBitQ estimator fields in SQ8.
Using each cluster centroid as the RaBitQ reference gives its distance-estimation factors and bound-error terms a narrow local range, enabling SQ8 encoding with little accuracy loss.
Conservative rounding of the error terms preserves valid pruning bounds.

\para{Page-efficient vector-level convergence.}
The full-precision vector file densely packs vectors in cluster order, allowing one page read to cover multiple candidates from the same cluster.
After the mid-precision gate, \sysname includes candidates on already selected pages at no extra read cost, then deduplicates page IDs and coalesces adjacent pages into larger reads.

\para{Inline and pooled re-ranking.}
The exact re-ranking step can run either inline in the query thread or in a dedicated thread pool.
In pooled mode, query threads pass finalized shortlists to dedicated re-ranking threads, which continuously issue new exact reads as earlier requests complete.
This continuous refill can drive SSD utilization closer to its IOPS limit.
Pooled re-ranking helps when I/O remains the bottleneck after distillation, but introduces handoff and queueing overhead.
Accordingly, \sysname uses pooled re-ranking for workloads that remain I/O-bound after the distill phase and inline re-ranking when the resulting I/O reduction from the distill phase shifts pressure toward computation.

\subsection{Discussion}\label{sec:design-discussion}

\sysname focuses on search optimization, while online-update support is orthogonal to its core design.
Prior work treats online updates as a dedicated design problem: SPFresh~\cite{SPFresh} supports updates in cluster-based disk-resident ANNS systems; FreshDiskANN~\cite{FreshDiskANN}, IP-DiskANN~\cite{IP-DiskANN}, and OdinANN~\cite{OdinANN} support updates in graph-based disk-resident ANNS systems.
\sysname could adapt SPFresh's local rebalancing: an insertion assigns a vector to its nearest cluster, stores its cluster-local quantized representation, and records its exact-vector location, while a deletion invalidates the vector's unique entry.
This approach would preserve the existing query workflow and simplify update handling because \sysname assigns each vector to exactly one cluster without replication, avoiding replica-version tracking and cleanup.

\section{Evaluation}
\label{sec:eval}

We evaluate \sysname against state-of-the-art disk-resident ANNS systems across diverse workloads and scales.
Our key findings include:
\begin{itemize}[leftmargin=*,nosep]
  \item At Recall@10=0.9, \sysname achieves on average $3.42\times$ PipeANN's throughput, 51.2\% lower latency, and 63.7\% less disk I/O across five datasets.
  \item \sysname has memory and storage footprints comparable to or, in most cases, smaller than those of existing disk-resident ANNS systems.
  \item \sysname progressively narrows query work; the saved I/O time outweighs distillation overhead, and its advantage grows with query concurrency.
\end{itemize}

\subsection{Experimental Setup}
\begin{table}[th]
  \centering
  \small
  \setlength{\tabcolsep}{1.5pt}
  \captionsetup{skip=5pt}
  \caption{\textbf{Datasets used in the evaluation.} \textit{\textbf{Dim.} is vector dimension, \textbf{Dist.} is distance metric for vector similarity.}}
  \label{tab:datasets}
  \begin{tabular*}{\columnwidth}{@{\extracolsep{\fill}}lrrcrrr@{}}
    \toprule
    \textbf{Dataset} & \textbf{Dim.} & \textbf{Type} & \textbf{Dist.} & \textbf{\#Vector} & \textbf{\#Query} & \textbf{Size(GiB)}\\
    \midrule
    DEEP-100M~\cite{Babenko_2016_DEEP_dataset} & 96 & float32 & L2 & 100M & 10K & 35.8 \\
    Text-to-Image~\cite{BigANN_Benchmark_neurips21} & 200 & float32 & IP & 100M & 100K & 74.5 \\
    Wiki~\cite{wikimedia_dumps} & 384 & float32 & L2 & 100M & 15.8K & 143.1 \\
    SIFT-1B~\cite{Jegou_2011_SIFT_dataset} & 128 & uint8 & L2 & 1B & 10K & 119.2 \\
    DEEP-1B~\cite{Babenko_2016_DEEP_dataset} & 96 & float32 & L2 & 1B & 10K & 357.6 \\
    \bottomrule
  \end{tabular*}
\end{table}

\para{Datasets.}
We comprehensively evaluate \sysname on five public datasets with diverse vector types and application domains, spanning scales from 100 million to 1 billion vectors, different dimensionalities, and different distance metrics (Table~\ref{tab:datasets}).
Specifically, we include popular datasets from the BigANN benchmark~\cite{BigANN_Benchmark_neurips21}, including DEEP-100M~\cite{Babenko_2016_DEEP_dataset} and the billion-scale SIFT-1B~\cite{Jegou_2011_SIFT_dataset} and DEEP-1B~\cite{Babenko_2016_DEEP_dataset} workloads.
We also include Yandex Text-to-Image~\cite{BigANN_Benchmark_neurips21}, a cross-modal dataset with base vectors derived from images and query vectors derived from text.
The base and query vectors follow different distributions in their shared embedding space, making it an out-of-distribution workload~\cite{RoarGraph_VLDB}.
Finally, following PipeRAG~\cite{PipeRAG} and existing work~\cite{DistVS}, we construct a 384-dimensional RAG-oriented dataset named Wiki using the BGE embedding model~\cite{BGE_Embedding}, with 100M base vectors generated from Wikipedia text data~\cite{wikimedia_dumps} and 15.8K query vectors generated from the MMLU text dataset~\cite{MMLU}.

\para{Baseline systems.}
We compare \textit{\sysname} with three state-of-the-art disk-resident ANNS systems: \textit{DiskANN}~\cite{DiskANN}, \textit{PipeANN}~\cite{pipeann_osdi}, and \textit{SPANN}~\cite{SPANN}.
\textit{DiskANN} and \textit{PipeANN} use PQ for their in-memory quantized vectors.
PQ is their default and recommended configuration and the setting under which they report strong performance.
Since \textit{\sysname} uses RaBitQ, we additionally include \textit{PipeANN}'s official RaBitQ implementation, denoted \textit{PipeANN+R}, configured with the same 2-bit-per-dimension RaBitQ representation as \sysname to isolate the effect of the quantization method.
Following \textit{PipeANN}~\cite{pipeann_osdi}, the \textit{DiskANN} and \textit{PipeANN} indexes use $R=96$ for graph degree, $L_{\mathrm{build}}=128$, and matching PQ widths (32 bytes for DEEP and SIFT, 50 for Text-to-Image, and 96 for Wiki).
At query time, \textit{DiskANN} uses a beam width of 4 to balance throughput and latency, while \textit{PipeANN} follows its default policy of dynamically increasing the I/O pipeline width from 4 to at most 32.
Since \textit{PipeANN} keeps a 1\% sampled entry graph in memory, it uses more DRAM than \textit{DiskANN}.
For a fair memory comparison with \textit{PipeANN}, we additionally include \textit{DiskANN+C}, a \textit{DiskANN} variant whose node cache is sized to approximately match \textit{PipeANN}'s memory usage.

\para{Testbed.}
All experiments run on a server equipped with two 48-core AMD EPYC 9455 processors, 384~GB of DRAM, and a Samsung 9100 PRO PCIe~5.0 NVMe SSD.
The SSD delivers up to 1 million IOPS for 4-KB random reads, as measured using fio~\cite{fio}.
The operating system is Ubuntu 24.04.4 with Linux 6.8.0.
Unless stated otherwise, experiments use all 96 physical cores and access the SSD through direct I/O.

\para{\sysname parameters.}
We set the target cluster size $T$ to 128 vectors.
\sysname uses 2-bit-per-dimension RaBitQ, aligning its per-vector quantization budget with \textit{PipeANN}~\cite{pipeann_osdi}, whose PQ representation uses approximately one byte per four dimensions.
We set the RaBitQ confidence parameter to its default value, $\epsilon_0=1.9$.
We jointly sweep the cluster-probe budget $P$ and the exact-vector budget $R$ and retain the best-performing configuration at each recall level to obtain \sysname's Pareto frontier over recall and performance.
Different $(P,R)$ combinations can achieve similar recall by shifting work between in-memory distillation and exact-vector I/O.
Empirically, \sysname favors smaller $P$ and larger $R$ at low concurrency (e.g., 1 thread), when SSD capacity remains ample, and larger $P$ and smaller $R$ at high concurrency (e.g., 96 threads), when the SSD becomes more stressed.
For each configuration, we set the early-termination window $W$ to approximately $0.3P$.
By default, \sysname assigns one eighth of its total threads to pooled re-ranking.
For the compute-intensive Wiki dataset and configurations with fewer than eight threads, each query thread performs inline re-ranking.

\para{Performance metrics.}
Our primary performance metrics are query throughput, measured in queries per second (QPS); query latency, measured as the average processing time per query; and disk I/O volume, measured as the average amount of data read per query.
We also report memory usage during query processing.
Following prior work~\cite{SPANN,DiskANN,Starling,pipeann_osdi}, we measure search quality using Recall@10 and report the average Recall@10 across all queries.

\subsection{Overall Performance}\label{sec:eval:overall}

\begin{figure*}[t]
  \centering
  \includegraphics[width=\textwidth]{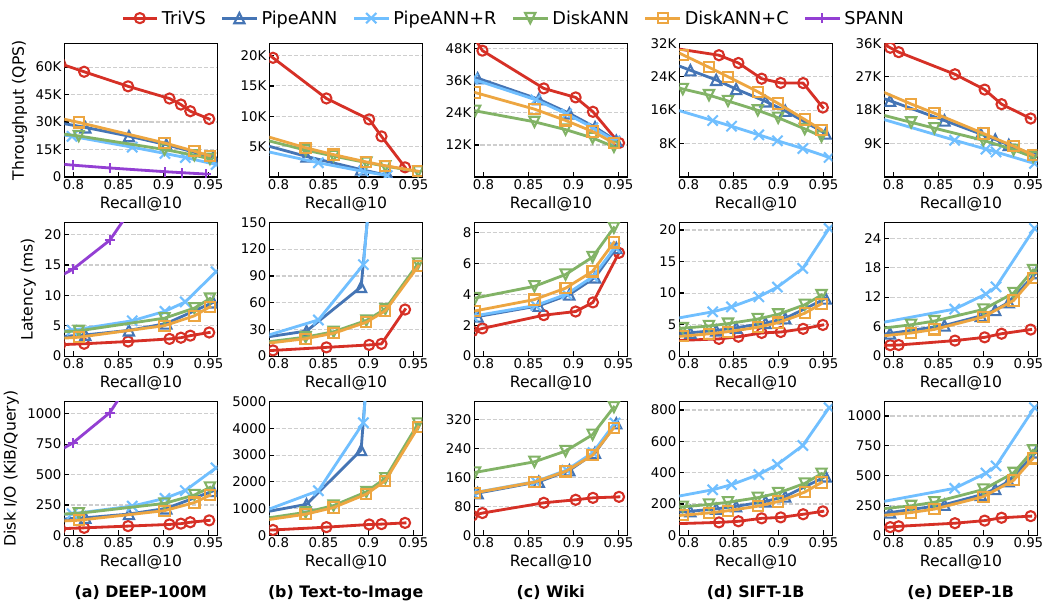}
  \captionsetup{skip=5pt}
  \caption{\textbf{Overall performance. Columns correspond to the five datasets; the three rows respectively report throughput, latency, and per-query disk I/O volume as Recall@10 varies.}
  \textit{\sysname improves throughput and latency by reducing disk I/O.}}
  \label{fig:overall-performance}
\end{figure*}

\para{Query throughput and latency.}
At Recall@10=0.9, the top and middle rows of Figure~\ref{fig:overall-performance} show that \sysname delivers the highest throughput and lowest latency on all five datasets.
Compared with \textit{PipeANN}, \sysname achieves 3.42$\times$ the throughput and 51.2\% lower latency on average.
Against the strongest baseline for each metric on each workload, \sysname delivers 1.29$\times$--3.94$\times$ the throughput (2.18$\times$ on average) and 28.5\%--68.1\% lower latency (44.8\% on average).
These gains hold across 100M- and billion-scale datasets, including the high-dimensional RAG-oriented Wiki and out-of-distribution Text-to-Image workloads.
Because \textit{SPANN} substantially trails the other baselines on DEEP-100M, our smallest workload, we omit it from the larger or higher-dimensional workloads, consistent with prior evaluations~\cite{Starling,pipeann_osdi,GPS_SIGMOD26}.

\para{I/O volume drives the gains.}
The bottom row of Figure~\ref{fig:overall-performance} shows that lower disk I/O drives these throughput and latency gains.
At Recall@10=0.9, \sysname reduces disk I/O by 47.3\%--74.7\% relative to the lowest-I/O baseline on each workload (57.7\% on average) and by 63.7\% relative to \textit{PipeANN} on average.
\sysname routes and selects candidates in memory, then reads only shortlisted exact vectors; graph-based baselines read graph records during approach, whereas \textit{SPANN} reads entire clusters during convergence.
A node cache lets \textit{DiskANN+C} avoid some approach-phase reads, but cannot eliminate them, leaving the same structural I/O bottleneck as \textit{PipeANN}.
RaBitQ alone does not explain these gains: despite using the same 2-bit-per-dimension representation as \sysname, \textit{PipeANN+R} performs more I/O and is slower than \textit{PipeANN} on all five workloads.
Its coarse reference centers cover dispersed graph vectors, whereas \sysname quantizes vectors within clusters.

\begin{table}[t]
  \centering
  \small
  \captionsetup{skip=5pt}
  \caption{\textbf{Query-serving memory and storage on DEEP-100M/1B (GiB).}
  \textit{SPANN is omitted on DEEP-1B because construction exceeds available memory.}}
  \label{tab:serving-footprint}
  \begin{tabular*}{0.8\columnwidth}{@{\extracolsep{\fill}}lrrrr@{}}
    \toprule
    & \multicolumn{2}{c}{\textbf{Memory}} & \multicolumn{2}{c}{\textbf{Storage}} \\[-2pt]
    \cmidrule(lr){2-3}\cmidrule(lr){4-5}
    \textbf{System} & \textbf{100M} & \textbf{1B} & \textbf{100M} & \textbf{1B} \\[-2pt]
    \midrule
    \textit{\sysname} & 4.14 & 42.0 & 42.98 & 430.0 \\
    \textit{PipeANN} & 5.91 & 43.4 & 80.20 & 798.2 \\
    \textit{PipeANN+R} & 6.27 & 59.3 & 81.41 & 814.1 \\
    \textit{DiskANN} & 3.02 & 29.9 & 79.31 & 792.8 \\
    \textit{DiskANN+C} & 6.09 & 43.5 & 79.31 & 792.8 \\
    \textit{SPANN} & 6.74 & -- & 235.01 & -- \\
    \bottomrule
  \end{tabular*}
\end{table}

\para{Memory usage.}
Table~\ref{tab:serving-footprint} shows that these performance gains do not come at the cost of higher query-serving memory.
At the 100M and 1B scales, \sysname uses 30\% and 3\% less memory than \textit{PipeANN}, respectively, and less than every other evaluated baseline except \textit{DiskANN}.
\sysname achieves this footprint with coarser clustering, which shrinks the centroid-routing index and frees DRAM for compact distillation codes.
Without a node cache, \textit{DiskANN} remains the lowest-memory option and can serve under tighter DRAM budgets, but at the cost of substantially lower throughput and higher latency.

\para{Storage usage.}
In addition, \sysname has the smallest storage footprint, using about 46\% less storage than graph-based baselines and 82\% less than \textit{SPANN}.
Compared with graph-based systems, \sysname avoids storing a per-vector adjacency graph.
Compared with \textit{SPANN}, \sysname assigns each vector to exactly one cluster, avoiding storage amplification from boundary replication.
\sysname preserves selective exact reads through in-memory distillation over coarser clusters.

\subsection{Micro Studies}

We next evaluate \sysname's key design choices and system behavior.
Unless otherwise noted, the micro studies use DEEP-100M at Recall@10=0.9.

\para{Each stage progressively reduces query work.}
Table~\ref{tab:progressive-work-reduction} quantifies how each stage narrows query work across all five datasets.
Early termination avoids scanning 8--33\% of the clusters probed during approach.
Mid precision evaluates only 0.8--4.9\% of the vectors screened at low precision, while exact re-ranking reads only 0.8--4.0\% of the mid-precision survivors.
Overall, \sysname reduces the candidate set by more than three orders of magnitude before exact-vector I/O.

\begin{table}[t]
  \centering
  \footnotesize
  \setlength{\tabcolsep}{3pt}
  \captionsetup{skip=4pt}
  \caption{\textbf{Progressive reduction in per-query work at Recall@10=0.9.}
  \textit{Low and Mid count vectors screened and refined at low and mid precision, respectively; Exact counts full-precision vectors read from disk for re-ranking. Probed/Scanned counts clusters selected during approach and scanned during distillation; early termination reduces the latter for easy queries.}}
  \label{tab:progressive-work-reduction}
  \begin{tabular*}{\columnwidth}{@{\extracolsep{\fill}}lrcrrr@{}}
    \toprule
    \textbf{Dataset} &
    \makecell{\textbf{Total}\\\textbf{\#Vector}} &
    \makecell{\textbf{Probed / Scanned}\\\textbf{\#Cluster}} &
    \makecell{\textbf{Low}\\\textbf{\#Vector}} &
    \makecell{\textbf{Mid}\\\textbf{\#Vector}} &
    \makecell{\textbf{Exact}\\\textbf{\#Vector}} \\
    \midrule
    DEEP-100M & 100M & 688.0 / 459.8 & 66.0K & 766.6 & 23.4 \\
    Text-to-Image & 100M & 3.07K / 2.82K & 273.6K & 13.4K & 112.8 \\
    Wiki & 100M & 712.0 / 573.1 & 84.8K & 641.2 & 25.4 \\
    SIFT-1B & 1B & 864.0 / 701.0 & 97.8K & 1.03K & 30.6 \\
    DEEP-1B & 1B & 768.0 / 621.1 & 89.4K & 1.22K & 32.4 \\
    \bottomrule
  \end{tabular*}
\end{table}

\begin{figure}[t]
  \centering
  \includegraphics[width=\columnwidth]{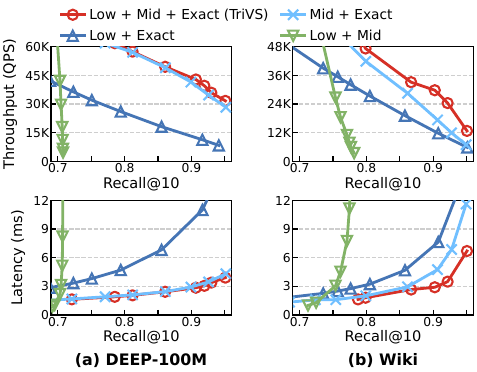}
  \captionsetup{skip=2pt}
  \caption{\textbf{Progressive-precision performance on DEEP-100M and Wiki.}
  \textit{Low + Exact requires larger exact shortlists because its estimates are coarse; Mid + Exact spends more computation by evaluating every scanned vector at mid precision, especially on high-dimensional Wiki; Low + Mid avoids exact I/O but cannot attain high recall.}}
  \label{fig:progressive-precision}
\end{figure}

\para{Progressive precision balances computation, I/O, and accuracy.}
The reductions in Table~\ref{tab:progressive-work-reduction} rely on applying higher precision only after the candidate set has narrowed.
Figure~\ref{fig:progressive-precision} compares the complete \textit{Low + Mid + Exact} pipeline with three two-level variants: \textit{Low + Exact}, \textit{Mid + Exact}, and \textit{Low + Mid}.
At Recall@10=0.9, the complete pipeline outperforms \textit{Mid + Exact} by 1.06$\times$ and 1.58$\times$, and \textit{Low + Exact} by 3.33$\times$ and 2.33$\times$ on DEEP-100M and Wiki, respectively.
\textit{Low + Exact} generates larger exact shortlists and more I/O because its estimates are coarse, whereas \textit{Mid + Exact} evaluates every scanned vector at mid precision, increasing computation, especially on Wiki.
\textit{Low + Mid} avoids exact-vector I/O, but without full-precision re-ranking it cannot attain high recall, plateauing at 0.71 and 0.78 Recall@10 on DEEP-100M and Wiki, respectively.
Thus, all three levels are necessary: low precision limits computation, mid precision limits I/O, and exact re-ranking enables high recall.

\begin{figure}[t]
  \centering
  \includegraphics[width=\columnwidth]{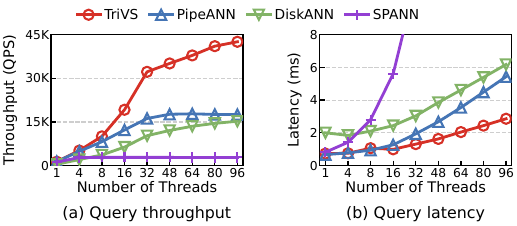}
  \captionsetup{skip=5pt}
  \caption{\textbf{Throughput and latency across thread counts.}
  \textit{\sysname remains close to PipeANN at one thread and delivers higher throughput and lower latency at high concurrency.}}
  \label{fig:deep100m-thread-scaling}
\end{figure}

\para{Scalability with query concurrency.}
Figure~\ref{fig:deep100m-thread-scaling} adds \sysname to the baseline scaling experiment in Figure~\ref{fig:baseline-qps-scaling} and compares throughput and latency across thread counts.
At one thread, \sysname remains close to PipeANN, with 5\% lower throughput and 10\% higher latency.
As concurrency grows and the baselines approach the SSD I/O limits identified in Figure~\ref{fig:baseline-qps-scaling}, \sysname's lower per-query I/O allows the same device capacity to serve more queries.
The change in \sysname's scaling slope around 32 threads partly reflects how it balances in-memory distillation and exact-vector I/O: at the same recall target, \sysname favors smaller $P$ and larger $R$ at lower concurrency, when SSD capacity is ample, and larger $P$ and smaller $R$ at higher concurrency, when the SSD is more stressed.
At 96 threads, \sysname delivers 2.43$\times$ and 2.79$\times$ the throughput of PipeANN and DiskANN, respectively, while reducing their latency by 47\% and 54\%.

\begin{figure}[t]
  \centering
  \includegraphics[width=\columnwidth]{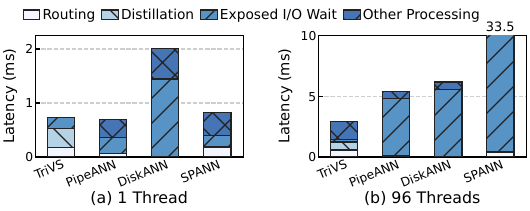}
  \captionsetup{skip=5pt}
  \caption{\textbf{Latency breakdown with 1 and 96 threads.}
  \textit{At high concurrency, \sysname's routing and in-memory distillation are offset by substantially less exposed I/O wait.}}
  \label{fig:deep100m-breakdown}
  \vspace{-5pt}
\end{figure}

\para{The saved I/O time outweighs distillation.}
Figure~\ref{fig:deep100m-breakdown} explains this scaling trend by comparing \sysname's latency components with PipeANN at one and 96 threads.
At one thread, \sysname's 0.53~ms of routing and distillation exceeds the 0.11~ms it saves in exposed I/O wait, leaving latency 7\% higher.
At 96 threads, however, \sysname spends 1.24~ms on routing and distillation but saves 4.53~ms of exposed I/O wait; this reduction outweighs its added in-memory work and lowers end-to-end latency by 46\%.

\begin{figure}[t]
  \centering
  \includegraphics[width=\columnwidth]{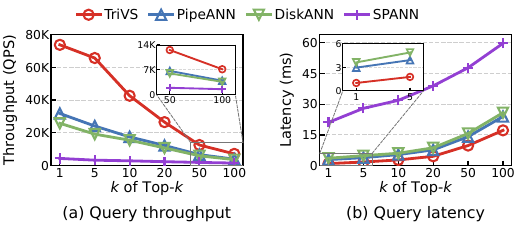}
  \captionsetup{skip=5pt}
  \caption{\textbf{Top-$k$ sensitivity on DEEP-100M at 0.9 recall.}
  \textit{\sysname maintains higher throughput and lower latency as the number of returned neighbors increases.}}
  \label{fig:topk-sensitivity}
\end{figure}

\para{Top-$k$ sensitivity.}
Figure~\ref{fig:topk-sensitivity} varies $k$, the number of nearest neighbors returned by ANNS, from 1 to 100 and maintains Recall@$k$=0.9.
\sysname delivers the highest throughput and lowest latency across the range, with $1.81\times$ PipeANN's throughput and 28\% lower latency at $k=100$.

\begin{figure}[t]
  \begin{minipage}[t]{0.49\columnwidth}
    \centering
    \includegraphics[width=\linewidth]{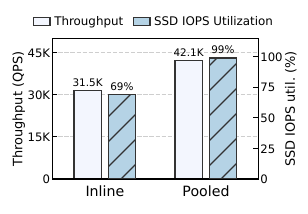}
    \captionsetup{skip=0pt}
    \captionof{figure}{\textbf{Inline and pooled re-ranking.}
    \textit{Pooling raises SSD utilization and query throughput.}}
    \label{fig:pooled-reranking}
  \end{minipage}
  \hfill
  \begin{minipage}[t]{0.49\columnwidth}
    \centering
    \includegraphics[width=\linewidth]{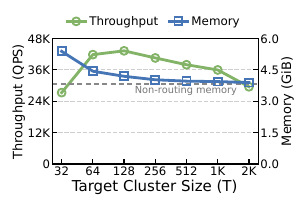}
    \captionsetup{skip=0pt}
    \captionof{figure}{\textbf{Sensitivity to target cluster size.}
    \textit{$T=128$ balances query throughput and memory usage.}}
    \label{fig:cluster-size}
  \end{minipage}
\end{figure}

\para{Pooled re-ranking.}
Figure~\ref{fig:pooled-reranking} compares inline and pooled re-ranking with the same 96-core budget.
By decoupling query execution from exact re-ranking, pooling steadily replenishes the I/O queue as reads complete, raising SSD IOPS utilization from 69\% to 99\% and throughput from 31.5K to 42.1K QPS.

\para{Sensitivity to cluster size.}
In Figure~\ref{fig:cluster-size}, we vary target cluster size $T$ and maintain the target Recall@10=0.9.
Our default $T=128$ balances query throughput and memory usage.
Smaller clusters increase centroid count and routing overhead, whereas larger clusters require more in-memory scanning and exact-vector reads to maintain recall.
Meanwhile, as $T$ grows, total memory approaches the measured 3.83~GiB non-routing component because compact per-vector quantization codes and other serving state do not shrink with the routing index.

\section{Related Work}

\para{Disk-resident ANNS systems.}
Cluster-based systems such as SPANN~\cite{SPANN} use an in-memory centroid index to select disk-resident clusters but fetch and scan each selected cluster in full.
Helmsman~\cite{Helmsman} augments SPANN with learned cluster pruning (not publicly released) and specialized SSD-array execution, but retains a single cluster granularity.

On the other hand, graph-based systems retain vector-level selectivity but read graph records from disk during proximity-graph traversal, as in DiskANN~\cite{DiskANN}.
Starling~\cite{Starling} and MARGO~\cite{MARGO_VLDB25} reorder graph records to improve page locality but require extra construction time and memory.
PageANN~\cite{PageANN_arxiv} builds a page-aligned graph, letting one read examine multiple nearby vectors.
These locality gains depend on multiple useful vectors sharing a page and diminish as higher dimensionality reduces page capacity~\cite{GPS_SIGMOD26}.
Cache-oriented designs include GoVector~\cite{GoVector_arxiv}, which combines hybrid node caching with similarity-aware reordering, and GPS~\cite{GPS_SIGMOD26}, which targets high-dimensional vectors by prioritizing adjacency lists over full vectors in memory and on disk.
PipeANN~\cite{pipeann_osdi} relaxes strict best-first-search ordering to pipeline dependent graph reads with computation, aligning graph search with SSD characteristics to achieve state-of-the-art performance.
Together, page-aware layouts, caching, and pipelined execution reduce graph I/O or hide its latency but cannot eliminate approach-phase I/O.
In contrast, \sysname eliminates this I/O through its in-memory approach and distill phases, reading only shortlisted full-precision vectors from disk during convergence.
Page-aware placement and caching are orthogonal and could further optimize \sysname's convergence-phase I/O; our evaluation shows that \sysname outperforms PipeANN despite its pipelined execution.

\para{Other ANNS systems.}
Conventional in-memory ANNS spans inverted-file~\cite{IVFADC,IVFADC+G+P}, graph-based~\cite{HNSW}, tree-based~\cite{Tree_ANNS}, hash-based~\cite{LSH}, and hybrid indexes~\cite{SPTAG}.
Beyond DRAM-only platforms, several ANNS systems build on emerging memory and storage technologies, including persistent memory~\cite{hmann_nips}, CXL-attached memory~\cite{CXL-ANNS}, processing-in-memory~\cite{PIMANN_ATC25}, and computational storage~\cite{SmartANNS}.
Accelerator-based systems use GPUs and FPGAs to exploit specialized parallelism for vector search~\cite{CAGRA_ICDE24,BANG,RUMMY_NSDI24,GustANN_SIGMOD,CMANNS_SIGMOD,FlowANN,FANNS_SC23,FusionANNS_FAST25}.
Distributed ANNS systems scale search across machines~\cite{Milvus_SIGMOD21,CoTra_SIGMOD26}, while disaggregated systems separate compute and memory~\cite{d-HNSW,DistVS}.
\sysname focuses on disk-resident vector search using the widely available single-node CPU, DRAM, and SSD platform.

\section{Conclusion}

We present \sysname, a new paradigm for disk-resident ANNS that decouples the approach and converge phases from a single search granularity through an in-memory distill phase.
\sysname organizes vector search as a trilogy that approaches coarsely, distills progressively, and converges exactly, combining the complementary I/O benefits of cluster and vector granularities to reduce per-query disk I/O.
Across diverse workloads, \sysname outperforms state-of-the-art systems in throughput and latency without increasing memory usage.




\normalem

\bibliographystyle{plain}
\bibliography{TriVS}

\end{document}